# Activation cross-section measurement of proton induced reactions on cerium

F. Tárkányi[1], A. Hermanne[2], F. Ditrói[1*], S. Takács[1], I. Spahn[3], S. Spellerberg[3]

1 Institute for Nuclear Research, Hungarian Academy of Sciences (ATOMKI), 4026 Debrecen, Hungary
2 Cyclotron Laboratory, Vrije Universiteit Brussel (VUB), Laarbeeklaan 103, 1090 Brussels, Belgium
3 Forschungszentrum Jülich GmbH, INM-5: Nuklearchemie, 52425 Jülich, Germany.

## Abstract

In the framework of a systematic study of proton induced nuclear reactions on lanthanides we have measured the excitation functions on natural cerium for the production of $^{142,139,138m,137}$Pr, $^{141,139,137m,137g,135}$Ce and $^{133}$La up to 65 MeV proton energy using the activation method with stacked-foil irradiation technique and high-resolution γ-ray spectrometry. The cross-sections of the investigated reactions were compared with the data retrieved from the TENDL-2014 and TENDL-2015 libraries, based on the latest version of the TALYS code system. No earlier experimental data were found in the literature.
The measured cross-section data are important for further improvement of nuclear reaction models and for practical applications in nuclear medicine, other labeling and activation studies.

[*] Corresponding author: ditroi@atomki.hu

## 1. Introduction

Radioisotopes of praseodymium and cerium could play an important role in radiotherapy and for visualizing biological processes. Among the recently investigated product radioisotopes, $^{142}$Pr (19.12 h), $^{141}$Ce (32.508 d) and $^{139}$Ce (137.641 d) [1-4] were proposed for internal radiotherapy. The rather short-lived $^{139}$Pr (4.41 h) has potential as a PET radiotracer [5] and $^{137m}$Ce as a SPECT radiotracer [6, 7]. Reliable and accurate knowledge of the excitation functions (shape and cross-section values over a wide energy range) is needed for selection of the optimal production routes and irradiation parameters. Until now, in most cases the quality of the theoretical predictions does not fulfill the requirements. In the literature no experimental data were found for production cross sections of these radioisotopes through proton induced reactions on cerium.

In the framework of our systematic study of the production possibilities of standard and emerging therapeutic radioisotopes of lanthanides via charged particle induced reactions, we already investigated earlier the deuteron induced production routes of these radioisotopes on neodymium [8], on praseodymium [9], on cerium [10] and proton and deuteron induced reactions on lanthanum [11, 12].

## 2. Experiment and data evaluation

The experimental techniques were similar to those reported in more detail in our previous works. Here we summarize only the most important factors for easier understanding and for compilation in nuclear reactions database. The main experimental parameters are summarized in Table 1.

Table 1. Experimental parameters

| **Reaction** | **$^{nat}Ce(p,x)$ (series 1)** | **$^{nat}Ce(p,x)$ (series 2)** |
|---|---|---|
| Incident particle | Proton | Proton |
| Method | Stacked foil | Stacked foil |
| Target and thickness (μm) | CeO (4.59-34.0, sedimented) | CeO (33.1-63.2, sedimented) |
| Target composition and thickness (μm) | Ti (10.9), Al (102), Tb (22.2), Ti (10.9), Al (102), Al (10), CeO (4.59-34.0), Al (100) Repeated 12 times | Tb (22.2), Al (377), La (25), Al(10), CeO (33.1-63.2), Al (100 ) Repeated 18 times |
| Accelerator | CGR-560 cyclotron of the Vrije Universiteit Brussel (VUB) in Brussels | Cyclone 90 cyclotron of the Université Catholique in Louvain la Neuve (LLN) |
| Primary energy (MeV) | 35 | 65 |
| Covered energy range (MeV) | 33.4-6.1 | 63.29-33.13 |
| Irradiation time (min) | 60 | 59 |
| Beam current (nA) | 100 | 90 |

The activation method with stacked-foil irradiation technique and off-line high-resolution γ-ray spectrometry was used. The target stacks were irradiated in a Faraday cup like target holder equipped with a secondary electron suppressor and a long collimator. The monitor foils were distributed uniformly in the stack and covered the whole energy range for re-measuring the excitation function of the monitor reaction and comparison with recommended values.

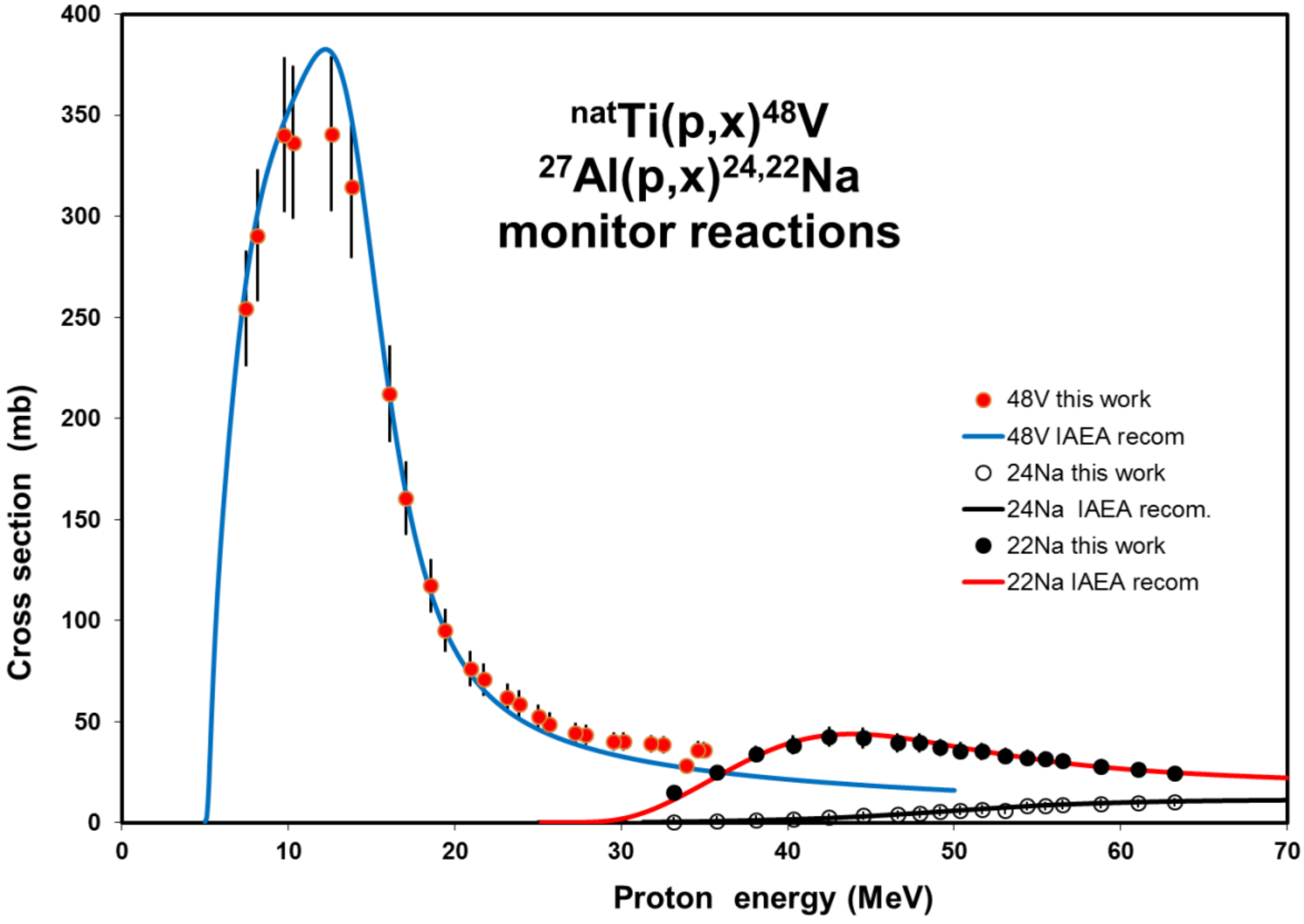

Fig. 1 The reproduced $^{nat}Ti(p,x)^{48}V$ and $^{27}Al(p,x)^{22,24}Na$ monitor reactions

The activity produced in the targets and monitor foils was measured non-destructively (without chemical separation) using high resolution HPGe gamma-ray spectrometers. Three series of measurements, at different detector–target distances, were done for all samples. Assessment of the foils of series 1 started at about 1.6-3.8 h, 21.5-25.1 h and 241.6-308.4 h after EOB, respectively, while for the samples of series 2 the cooling times were 5.8-8.4 h, 28.2-33.3 h and 220.4-296.1 after EOB, respectively. The decay data used were taken from the online database NuDat2 [13] and are presented in Table 2.

Effective beam intensities and the energy scale were determined by using the excitation functions of the $^{24}Al(p,x)^{22,24}Na$ and $^{nat}Ti(p,x)^{48}V$ reactions [14], simultaneously re-measured over the whole energy range. For illustration, the excitation functions of these monitor reactions, in comparison with the IAEA recommended data are shown in Fig. 1 and they are good agreement. Reaction Q-values (see Table 2) were taken from the BNL Q-value calculator [15]. The preliminary determination of beam energy was made by calculation using stopping powers of Andersen and Ziegler [16]. The final energy scale was adjusted to the fitted monitor reaction [17].

For estimation of the uncertainty of the effective beam energy in the target foils, cumulative effects of possible uncertainties in primary energy and target thickness were taken into account together with the effect of the energy straggling and of the correction for the monitor reaction.

Elemental cross-sections were deduced using the total numbers of cerium atoms in the target. The uncertainties of cross sections were obtained from the sum in quadrature of all individual contributions (beam current (7%), beam-loss corrections (maximum 1.5%), target thickness (3 %), detector efficiency (5%), photo peak area determination and counting statistics 1-20 %) [18].

The uncertainty of the non-linear contributing processes like irradiation, cooling and measuring time, and half-life was not considered.

Table 2. Decay characteristics of the investigated activation products and Q-values of contributing reactions

| **Nuclide**<br>**Decay path**<br>***Level(keV)*** | **Half-life** | **$E_\gamma$ (keV)** | **$I_\gamma$ (%)** | **Contributing reaction** | **Q-value (keV)** |
|---|---|---|---|---|---|
| **$^{142}$Pr**<br>ε: 0.0164%<br>$\beta^-$: 99.9836 % | 19.12 h | 1575.6 | 3.7 | $^{142}$Ce(p,n) | -1526.83 |
| **$^{139}$Pr**<br>ε: 100 % | 4.41 h | 255.11<br>1347.33<br>1630.67 | 0.236<br>0.473<br>0.343 | $^{140}$Ce(p,2n)<br>$^{142}$Ce(p,4n) | -12111.71<br>-24707.91 |
| **$^{138m}$Pr**<br>ε: 100 %<br>*364* | 2.12 h | 302.7<br>390.9<br>547.5<br>788.7<br>1037.8 | 80<br>6.1<br>5.2<br>100<br>100 | $^{138}$Ce(p,n)<br>$^{140}$Ce(p,3n)<br>$^{142}$Ce(p,5n) | -5219.3<br>-21872.4<br>-34468.6 |
| **$^{137}$Pr**<br>ε: 100 % | 1.28 h | 160.32<br>433.89<br>836.65 | 0.97<br>1.28<br>1.78 | $^{138}$Ce(p,2n)<br>$^{140}$Ce(p,4n)<br>$^{142}$Ce(p,6n) | -13220.6<br>-29873.66<br>-42469.85 |
| **$^{141}$Ce**<br>$\beta^-$: 100 % | 32.508 d | 145.44 | 48.29 | $^{142}$Ce(p,pn)<br>$^{141}$La decay | -7168.05<br>-8890.52 |
| **$^{139}$Ce**<br>ε: 100 % | 137.641 d | 165.86 | 80 | $^{140}$Ce(p,pn)<br>$^{142}$Ce(p,p3n)<br>$^{139}$Pr decay | -9200.29<br>-21796.49<br>-12111.71 |
| **$^{137m}$Ce**<br>IT: 99.21 %<br>ε: 0.79 %<br>*254.295* | 34.4 h | 169.26<br>824.82<br>254.29 | 0.45<br>0.450<br>0.111 | $^{138}$Ce(p,pn)<br>$^{140}$Ce(p,p3n)<br>$^{142}$Ce(p,p5n) | -9721.25<br>-26374.34<br>-38970.53 |
| **$^{137g}$Ce** | 9.0 h | 447.15 | 1.68 | $^{138}$Ce(p,pn) | -9721.25 |
| ε: 100 % | | | | $^{140}$Ce(p,p3n)<br>$^{142}$Ce(p,p5n)<br>$^{137m}$Pr decay | -26374.34<br>-38970.53<br>-9721.25 |
| **$^{135}$Ce**<br>ε: 100 % | 17.7 h | 265.56<br>300.07<br>518.05<br>572.26<br>606.76<br>783.59 | 41.8<br>23.5<br>13.6<br>10.4<br>18.8<br>10.6 | $^{136}$Ce(p,pn)<br>$^{138}$Ce(p,p3n)<br>$^{140}$Ce(p,p5n)<br>$^{142}$Ce(p,p7n)<br>$^{135}$Pr decay | -9963.5<br>-27166.3<br>-43819.4<br>-56415.6<br>-14426.2 |
| **$^{133}$La**<br>ε: 100 % | 3.912 h | 278.84<br>302.38 | 2.44<br>1.61 | $^{136}$Ce(p,2p2n)<br>$^{138}$Ce(p,2p4n)<br>$^{140}$Ce(p,2p6n)<br>$^{133}$Ce decay | -24445.6<br>-41650.8<br>-58298.6<br>-28304.1 |

When complex particles are emitted instead of individual protons and neutrons the Q-values have to be decreased by the respective binding energies of the compound particles: np-d, +2.2 MeV; 2np-t, +8.48 MeV; n2p-$^{3}$He, +7.72 MeV; 2n2p-α, +28.30 MeV.

Isotopic abundances: $^{136}$Ce (0.19 %), $^{138}$Ce (0.25 %), $^{140}$Ce (88.48 %), $^{142}$Ce (11.08 %),

## 3. Comparison with the theoretical results in TENDL libraries

The experimental data were compared with cross section data reported in the TENDL-2014 [19], and TENDL-2015 [20] databases, calculated using the TALYS code [21] with global input parameters. In this work the results from the both latest libraries are given in order to follow the development of the TENDL database.

The reaction cross sections of an investigated radionuclide on individual stable target isotopes were summed, weighted with target isotope’s abundance in natural cerium, allowing comparison with elemental cross-sections and understanding the importance of the contributing processes. The first experimental data presented here allow to test the prediction capabilities of the TALYS code.

# 4. Results

## *4.1 Excitation functions*

We deduced excitation functions for the $^{142,139,138m,137}Pr$, $^{141,139,137m,137,135}Ce$ and $^{140}La$ gamma emitter activation products. The excitation functions are shown in Fig. 2-11 while the numerical data are collected in Tables 3 and 4. The element cerium has 4 naturally occurring stable isotopes. The contributing reactions for the investigated products are shown in Table 2.

Table 3. Experimental cross-sections for the $^{nat}Ce(p,x)^{142,139,138m,137}Pr$, $^{141}Ce$ reactions

| | | $^{142}Pr$ | | $^{139}Pr$ | | $^{138m}Pr$ | | $^{137}Pr$ | | $^{141}Ce$ | |
|---|---|---|---|---|---|---|---|---|---|---|---|
| E | ΔE | σ | Δ σ | σ | Δ σ | σ | Δ σ | σ | Δ σ | σ | Δ σ |
| MeV | | mb | | | | | | | | | |
| **Series 1** | | | | | | | | | | | |
| 33.16 | 0.2 | | | 303.11 | 60.37 | 316.35 | 38.37 | 66.23 | 28.82 | 17.30 | 1.95 |
| 30.99 | 0.28 | | | 353.34 | 65.09 | 241.21 | 30.11 | | | 13.62 | 1.54 |
| 29.07 | 0.34 | 1.62 | 0.36 | 312.92 | 46.63 | 255.55 | 30.03 | | | 18.28 | 2.05 |
| 26.81 | 0.42 | 2.71 | 0.60 | 315.75 | 42.28 | 104.92 | 12.59 | | | 15.09 | 1.70 |
| 24.92 | 0.49 | | | 467.72 | 61.26 | 22.55 | 3.42 | | | 14.82 | 1.70 |
| 23.06 | 0.55 | 1.04 | 0.35 | 388.02 | 44.60 | | | | | 10.14 | 1.14 |
| 20.96 | 0.63 | 1.40 | 0.49 | 355.23 | 43.95 | | | | | 7.20 | 0.82 |
| 18.75 | 0.70 | | | 346.23 | 46.86 | | | | | 4.84 | 0.58 |
| 16.10 | 0.80 | 2.83 | 0.64 | 177.62 | 22.86 | | | | | 1.62 | 0.19 |
| 12.86 | 0.91 | 7.55 | 2.55 | | | | | | | 0.13 | 0.07 |
| 11.02 | 0.97 | 10.32 | 2.23 | | | | | | | | |
| 8.38 | 1.07 | 5.17 | 0.96 | | | | | | | | |
| **Series 2** | | | | | | | | | | | |
| 63.20 | 0.2 | | | | | 32.54 | 3.69 | 144.43 | 95.31 | 11.84 | 1.35 |
| 61.06 | 0.27 | | | | | 38.64 | 4.35 | | | 13.62 | 1.53 |
| 58.81 | 0.35 | | | | | 33.80 | 3.80 | 158.64 | 31.31 | 10.39 | 1.17 |
| 56.51 | 0.43 | | | | | 46.36 | 5.21 | 202.41 | 52.58 | 13.28 | 1.50 |
| 55.45 | 0.47 | | | | | 42.78 | 4.81 | 199.71 | 37.28 | 11.36 | 1.29 |
| 54.38 | 0.51 | | | | | 48.48 | 5.46 | 273.22 | 63.39 | 12.90 | 1.46 |
| 53.07 | 0.55 | | | | | 58.22 | 6.54 | 350.01 | 47.78 | 16.83 | 1.89 |
| 51.65 | 0.60 | | | | | 41.11 | 4.62 | 246.05 | 34.10 | 10.18 | 1.15 |
| 50.36 | 0.65 | | | | | 51.97 | 5.84 | 314.73 | 48.43 | 11.87 | 1.34 |
| 49.16 | 0.69 | | | | | 57.84 | 6.52 | 356.10 | 92.79 | 12.90 | 1.46 |
| 47.91 | 0.74 | | | | | 69.56 | 7.82 | 581.71 | 94.80 | 15.55 | 1.75 |
| 46.58 | 0.78 | | | | | 57.02 | 6.41 | 494.60 | 63.65 | 12.40 | 1.40 |
| 44.53 | 0.85 | | | | | 72.90 | 8.23 | 629.24 | 136.74 | 13.84 | 1.58 |
| 42.48 | 0.93 | | | | | 85.23 | 9.61 | 429.07 | 94.52 | 13.69 | 1.57 |
| 40.34 | 1.00 | | | 134.47 | 27.84 | 94.79 | 10.65 | 393.57 | 61.71 | 11.93 | 1.36 |
| 38.09 | 1.08 | | | 179.05 | 43.81 | 134.81 | 15.17 | 300.00 | 117.58 | 11.77 | 1.37 |
| 35.72 | 1.16 | | | 197.71 | 35.01 | 226.81 | 25.48 | 219.31 | 66.46 | 15.51 | 1.76 |
| 33.14 | 1.25 | | | 188.15 | 42.33 | 217.42 | 24.42 | 9.56 | 44.17 | 13.62 | 1.53 |

Table 4. Experimental cross-sections for $^{nat}Ce(p,x)^{139,137m,137g,135}Ce$ and $^{133}La$ reactions

| | | $^{139}Ce$ | | $^{137m}Ce$ | | $^{137g}Ce$ | | $^{135}Ce$ | | $^{133}La$ | |
|---|---|---|---|---|---|---|---|---|---|---|---|
| **E** | **ΔE** | σ | **Δ** σ | σ | **Δ** σ | σ | **Δ** σ | σ | **Δ** σ | σ | **Δ** σ |
| **MeV** | | **mb** | | | | | | | | | |
| **Series 1** | | | | | | | | | | | |
| 33.16 | 0.20 | 413.05 | 46.38 | 2.62 | 0.54 | | | 1.25 | 0.17 | | |
| 30.99 | 0.28 | 343.49 | 38.57 | 1.63 | 0.38 | | | 1.19 | 0.14 | | |
| 29.07 | 0.34 | 541.58 | 60.80 | 1.90 | 0.35 | | | 2.15 | 0.25 | | |
| 26.81 | 0.42 | 704.01 | 79.03 | 1.82 | 0.33 | | | 2.14 | 0.25 | | |
| 24.92 | 0.49 | 772.77 | 86.78 | 2.75 | 0.60 | | | 2.07 | 0.28 | | |
| 23.06 | 0.55 | 842.17 | 94.53 | 2.96 | 0.51 | | | 1.87 | 0.27 | | |
| 20.96 | 0.63 | 736.47 | 82.67 | 2.35 | 0.45 | | | 1.17 | 0.23 | | |
| 18.75 | 0.70 | 698.22 | 78.39 | 1.32 | 0.48 | | | 1.14 | 0.25 | | |
| 16.1 | 0.80 | 387.10 | 43.45 | 1.18 | 0.23 | | | | | | |
| 12.86 | 0.91 | 29.26 | 3.30 | | | | | | | | |
| 11.02 | 0.97 | 0.84 | 0.39 | | | | | | | | |
| 8.38 | 1.07 | 0.22 | 0.04 | | | | | | | | |
| **Series 2** | | | | | | | | | | | |
| 63.2 | 0.20 | 182.05 | 20.46 | 59.93 | 8.08 | 173.13 | 22.83 | 101.39 | 11.44 | 15.24 | 6.15 |
| 61.06 | 0.27 | 209.25 | 23.49 | 70.65 | 8.48 | 188.11 | 23.26 | 70.00 | 7.89 | 16.48 | 4.50 |
| 58.81 | 0.35 | 163.80 | 18.40 | 58.59 | 7.01 | 147.25 | 17.65 | 26.50 | 3.01 | 12.42 | 3.51 |
| 56.51 | 0.43 | 212.75 | 23.89 | 68.02 | 8.45 | 225.45 | 27.58 | 12.00 | 1.48 | 6.01 | 4.63 |
| 55.45 | 0.47 | 184.53 | 20.73 | 48.83 | 6.48 | 200.16 | 24.72 | 5.30 | 0.79 | | |
| 54.38 | 0.51 | 208.59 | 23.43 | 63.08 | 7.70 | 245.37 | 29.07 | | | 4.98 | 3.60 |
| 53.07 | 0.55 | 283.55 | 31.83 | 72.04 | 8.27 | 306.00 | 34.78 | 2.93 | 0.42 | 4.30 | 1.85 |
| 51.65 | 0.60 | 173.09 | 19.43 | 43.31 | 5.29 | 223.51 | 25.83 | 1.75 | 0.33 | 3.07 | 1.77 |
| 50.36 | 0.65 | 205.30 | 23.05 | 49.92 | 6.13 | 315.00 | 36.31 | 2.19 | 0.46 | | |
| 49.16 | 0.69 | 220.03 | 24.71 | 57.10 | 6.99 | 366.54 | 42.63 | 1.73 | 0.49 | | |
| 47.91 | 0.74 | 282.04 | 31.66 | 61.46 | 7.25 | 510.74 | 57.98 | 1.26 | 1.27 | | |
| 46.58 | 0.78 | 232.27 | 26.08 | 45.07 | 5.79 | 373.49 | 42.86 | 1.66 | 0.41 | | |
| 44.53 | 0.85 | 265.18 | 29.79 | 44.36 | 6.37 | 491.39 | 57.55 | 2.01 | 0.67 | | |
| 42.48 | 0.93 | 279.70 | 31.43 | 31.46 | 5.50 | 466.00 | 55.78 | 1.29 | 0.68 | | |
| 40.34 | 1.00 | 259.37 | 29.13 | 16.51 | 3.70 | 349.51 | 41.56 | | | | |
| 38.09 | 1.08 | 267.49 | 30.07 | | | 242.71 | 31.62 | | | | |
| 35.72 | 1.16 | 351.17 | 39.44 | 6.80 | 3.23 | 210.41 | 27.82 | | | | |
| 33.14 | 1.25 | 309.47 | 34.74 | 4.97 | 2.54 | 54.77 | 11.04 | | | | |

### 4.1.1 $^{142}$Pr (direct)

As seen in Fig. 2 only a few data points were obtained in the low energy irradiation and due to the low counting statistics the data are scattered. The ground state of $^{142}$Pr ($T_{1/2}$ = 19.12 h) is produced directly via the $^{142}$Ce(p,n) reaction and via isomeric decay of the 14.6 min half-life metastable state. The TENDL library (quasi identical) results overestimate the experimental data by a factor of nearly two above 12 MeV.

### 4.1.2 $^{139}$Pr (direct)

The experimental and theoretical data for production of $^{139}$Pr ($T_{1/2}$ = 4.41 h) are shown in Fig. 3. The TENDL library (identical) results overestimate the experimental data by a factor of more than two.

### 4.1.3 $^{138m}$Pr (direct)

The radionuclide $^{138}$Pr has two observable isomeric states. The 1.45 min half-life ground state and the 2.12 h half-life higher laying isomeric states are decaying independently by EC or $\beta^+$ to stable $^{138}$Ce. The excitation function of the longer-lived $^{138m}$Pr is shown in Fig. 4. The TENDL library (TENDL-2015 is 20% higher than TENDL-2014) results overestimate strongly the experimental data.

### 4.1.4 $^{137}$Pr (direct)

Due to the relatively long cooling time after EOB, the gamma-lines of $^{137}$Pr ($T_{1/2}$=1.28 h) were detected with poor statistics and the deduced cross-section data are more scattered (Fig. 5). The TENDL library (quasi identical) results are slightly energy shifted against the experimental data but are representing them well.

#### 4.1.5 $^{141}$Ce (cumulative)

The radionuclide $^{141}$Ce ($T_{1/2}$ = 32.508 d) is produced directly and from the decay of shorter-lived $^{141}$La ($T_{1/2}$ = 3.92 h). The experimental data shown in Fig. 6 are cumulative as they were obtained from spectra measured after "complete" decay of $^{141}$La. The TENDL library (quasi identical) results overestimate the experimental data above 30 MeV.

#### 4.1.6 $^{139}$Ce (cumulative)

The cumulative cross sections for production of the long-lived $^{139}$Ce ($T_{1/2}$ = 137.641 d), containing contribution of parent $^{139}$Pr ($T_{1/2}$ = 4.41 h) after its "complete" decay, are shown in Fig. 7. The TENDL library (identical) results for cumulative formation slightly overestimate the experimental data and show the importance of the parent decay.

#### 4.1.7 $^{137m}$Ce (direct)

The radionuclide $^{137}$Ce has isomeric states: the higher lying $^{137m}$Ce with a half-life of 34.4 h and the shorter-lived ground state $^{137g}$Ce ($T_{1/2}$ = 9.0 h). We can deduce cross-section data for both states. Decay of $^{137}$Pr ($T_{1/2}$ = 1.28 h) does not contribute to formation of $^{137m}$Ce (Fig. 8). The TENDL library (quasi identical) results overestimate the experimental data by a factor of three near the maximum.

#### 4.1.8 $^{137g}$Ce (cum)

For formation of the $^{137g}$Ce (9.0 h half-life) ground state contributions from the decay of the longer-lived isomeric state ($T_{1/2}$ = 34.4 h, IT: 99.21 %) and from the decay of $^{137}$Pr ($T_{1/2}$ = 1.28 h), in addition to direct production, exist. We deduced the cross-sections from our first series of spectra, measured in the time interval 5.8-8.4 h after EOB, the small contribution from the decay of the metastable isomer was subtracted (long half-life and the relatively small cross-sections of the isomeric state). The cross-sections shown in the Fig. 9 represent the sum of direct production and the small contribution of the total decay of parent $^{137}$Pr. The TENDL library results for cumulative formation are somewhat energy shifted and slightly overestimate the experimental data.

#### *4.1.9 $^{135}$Ce (cumulative)*

The cumulative cross-section of the ground state of $^{135}$Ce ($T_{1/2}$ = 17.7 h) includes the contributions from the total decays of the short-lived isomeric state $^{135m}$Ce ($T_{1/2}$ = 20 s, IT: 100 %) and from parent $^{135}$Pr ($T_{1/2}$ = 24 min)(Fig. 10). The TENDL library (quasi identical for direct, slight energy shifted for the cumulative) results overestimate the experimental data by a factor of two.

#### *4.1.10 $^{133}$La (cumulative)*

The radionuclide $^{133}$La is produced directly (mostly via (p,$\alpha$xn) reactions) and through the $^{133}$Pr-decay chain. The $^{133}$Pr decays to the $^{133}$Ce ($T_{1/2}$ = 97 min) ground state. In the first series of spectra even the high intensity gamma-lines of the longer-lived, high spin isomeric states of $^{133}$Ce ($T_{1/2}$ = 5.1 h) were not observed, indicating the very low contribution to the production of $^{133}$La. In such a way the measured cross-section is cumulative containing practically direct production and the contribution from the $^{133}$Pr-$^{133g}$Ce decay chain. The TENDL libraries (TENDL-2014 is lower at the maximum than TENDL-2015) results underestimate the experimental data above 60 MeV.

## 5. Integral yields

Integrated yields for a given incident energy down to the reaction threshold as a function of the bombarding energy were calculated from fitted curves to our experimental cross-section data. The results for physical yields (Bonardi et al. [22] ) are presented in Figs. 12-13 in comparison with the experimental yield data of Dmitriev and Molin for production of $^{139}$Ce [23]. The literature data is somewhat lower than our new values at the single 22 MeV energy point.

## 6. Production routes of medically relevant radioisotopes

Among the radioisotopes discussed in this work, $^{142}$Pr, $^{141}$Ce and $^{139}$Ce are potentially important for internal radiotherapy, while $^{139}$Pr has potential as a PET- tracer.

The possible production routes were discussed in detail in our previous paper [10]. The significance of the presently investigated proton induced reactions on cerium are the favorable conditions for direct production of $^{142}$Pr and $^{139}$Pr and indirect production of $^{139}$Ce:

- For $^{142}$Pr ($T_{1/2}$ = 19.12 h) production, the (p,n) and (d,2n) charged particle reactions produce a high specific activity, no-carrier added product. Although the production yield for the (d,2n) reaction is significantly higher compared to the (p,n) reaction, deuteron induced reactions require higher energy cyclotrons.
- Out of the $^{140}$Ce(p,2n), $^{140}$Ce(d,3n), $^{139}$La($\alpha$,2n) main production routes for $^{139}$Pr ($T_{1/2}$ = 4.41 h ) the first two are the most productive, but they require highly enriched targets. The (d,3n) requires higher energy machines and the alpha beam is rare at modern cyclotrons.
- A broad range of charged particle reactions for production of $^{139}$Ce ($T_{1/2}$ = 137.641 d) exists. The indirect routes through the decay of $^{139}$Pr ($T_{1/2}$ = 4.41 h), based on the more productive $^{140}$Ce(p,2n) $^{139}$Pr$\rightarrow$$^{139}$Ce reaction and the $^{141}$Pr(p,p2n) $^{139}$Pr$\rightarrow$$^{139}$Ce reaction are the most interesting (irradiation in saturation and chemical separation of $^{139}$Pr).

## 7. Summary

We have measured the excitation functions on natural cerium for production of $^{142,139,138m,137}$Pr, $^{141,139,137m,137g,135}$Ce and $^{133}$La up to 65 MeV and presented cross section results for the first time in the whole energy range. The experimental cross sections are compared with the TALYS theoretical calculation, showing systematic disagreements. The differences between TENDL-2014 and TENDL-2015 are non-systematic. The proton induced reactions on cerium for the medically relevant $^{142,139}$Pr and $^{139}$Ce seem to be potentially important.

***Acknowledgements***

This work was performed in the frame of the HAS-FWO Vlaanderen (Hungary-Belgium) project. The authors acknowledge the support of the research project and of the respective institutions (VUB, LLN) in providing the beam time and experimental facilities.

**Figures (from2.)**

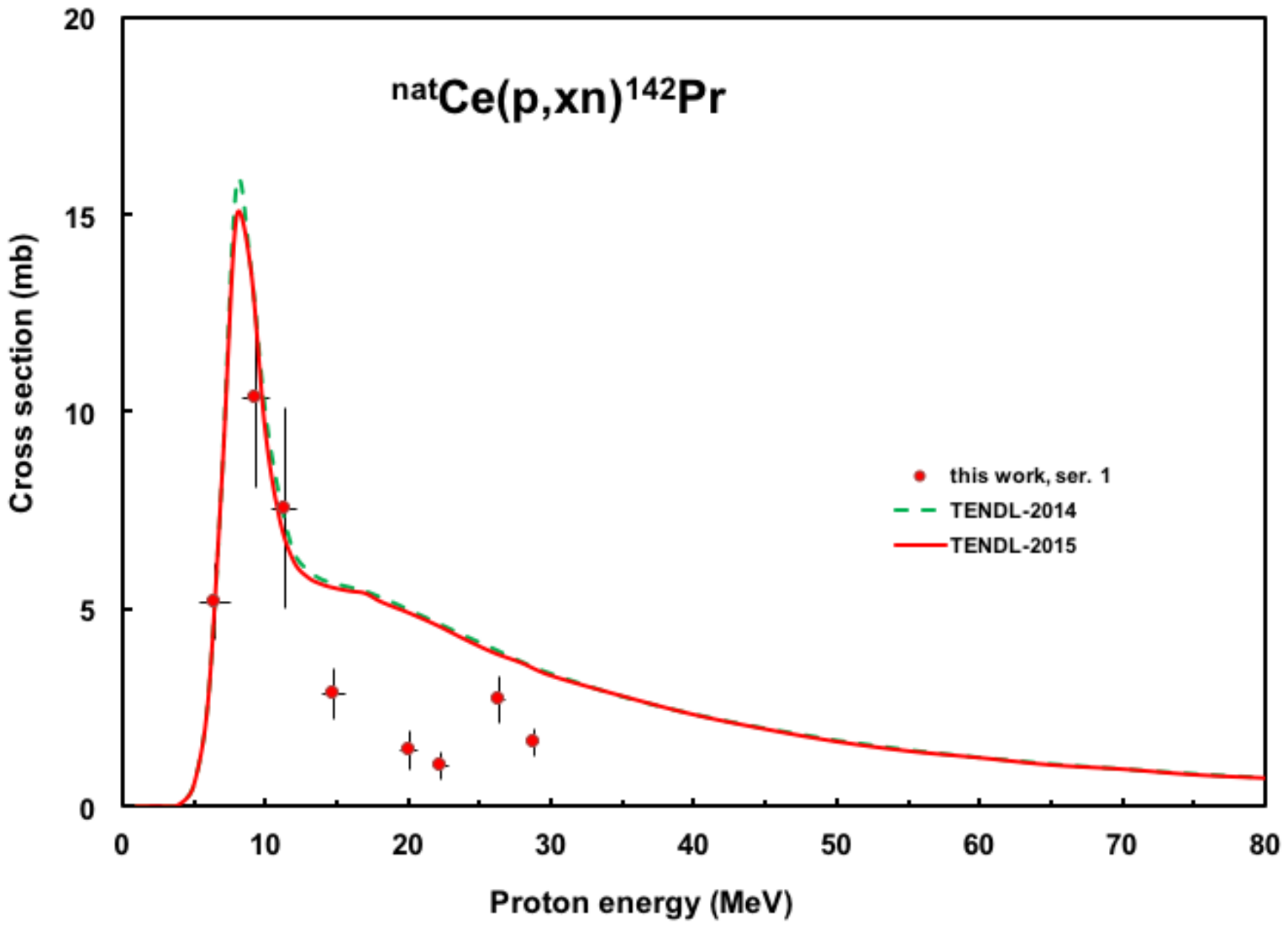

Fig. 2. Experimental and theoretical excitation functions for the $^{nat}Ce(p,xn)^{142}Pr$ reaction

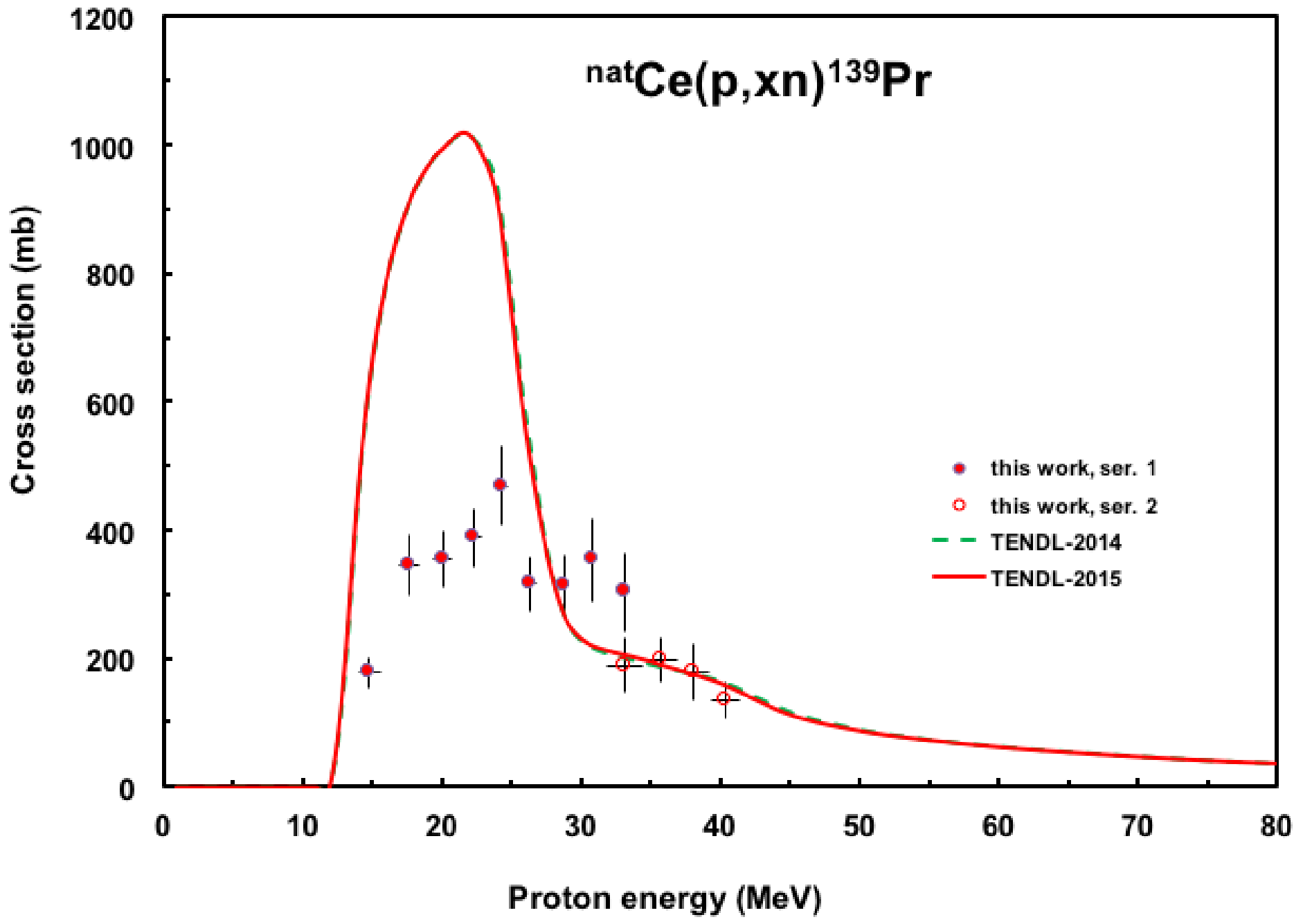

Fig. 3. Experimental and theoretical excitation functions for the $^{nat}Ce(p,xn)^{139}Pr$ reaction

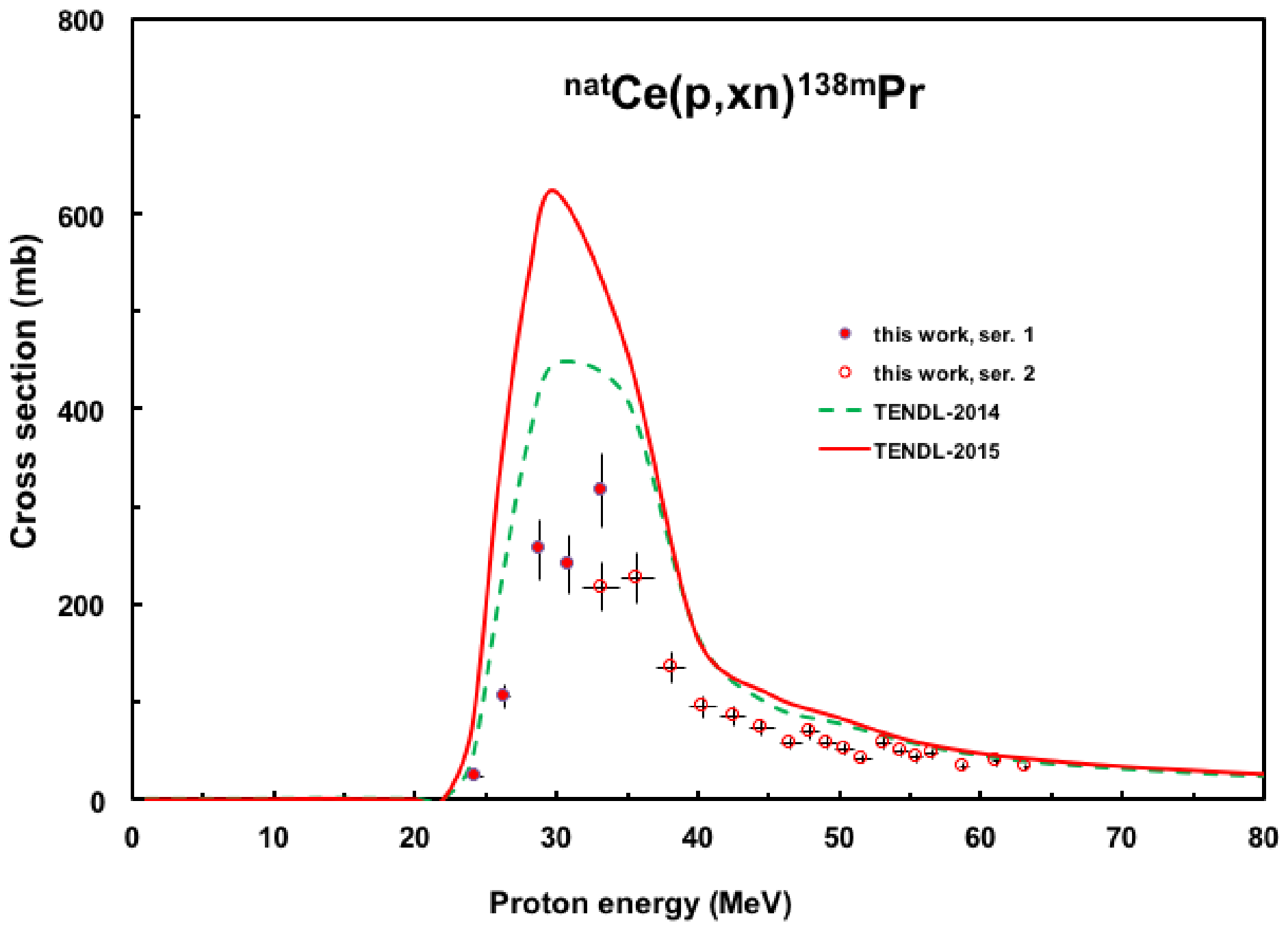

Fig. 4. Experimental and theoretical excitation functions for the $^{nat}Ce(p,xn)^{138m}Pr$ reaction

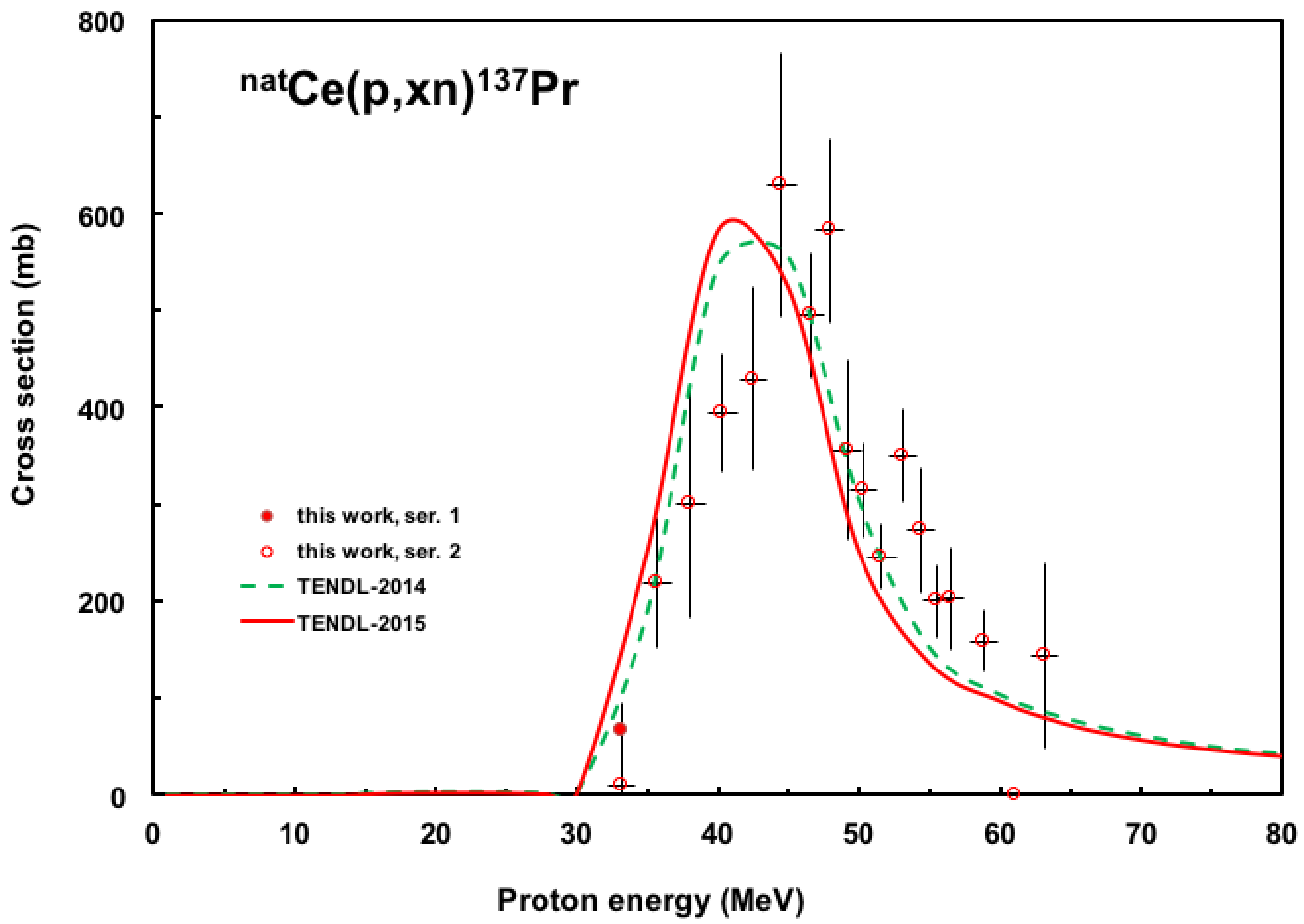

Fig. 5. Experimental and theoretical excitation functions for the $^{nat}Ce(p,xn)^{137}Pr$ reaction

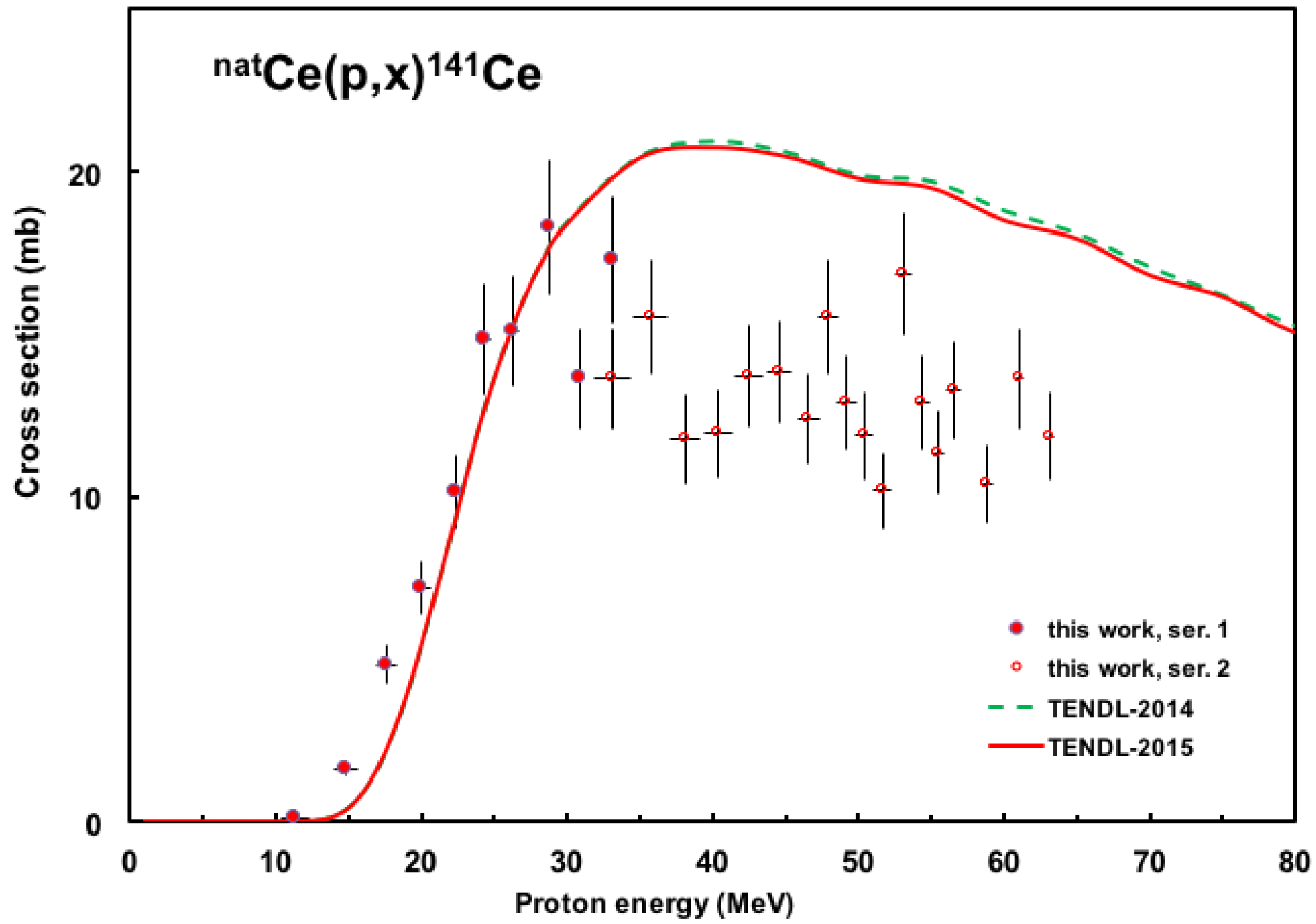

Fig. 6. Experimental and theoretical excitation functions for the $^{nat}Ce(p,x)^{141}Ce$ reaction

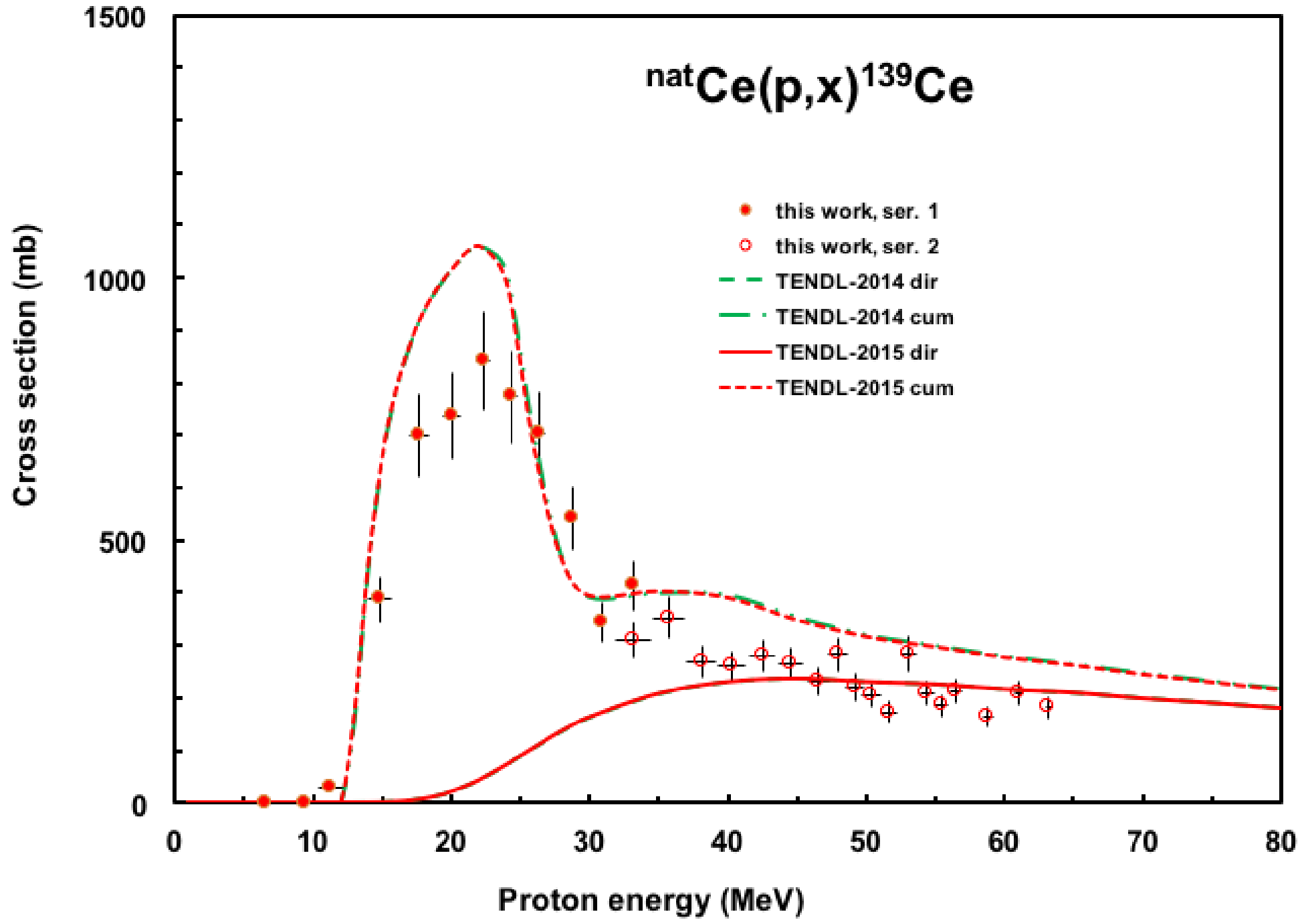

Fig. 7. Experimental and theoretical excitation functions for the $^{nat}Ce(p,x)^{139}Ce$ reaction

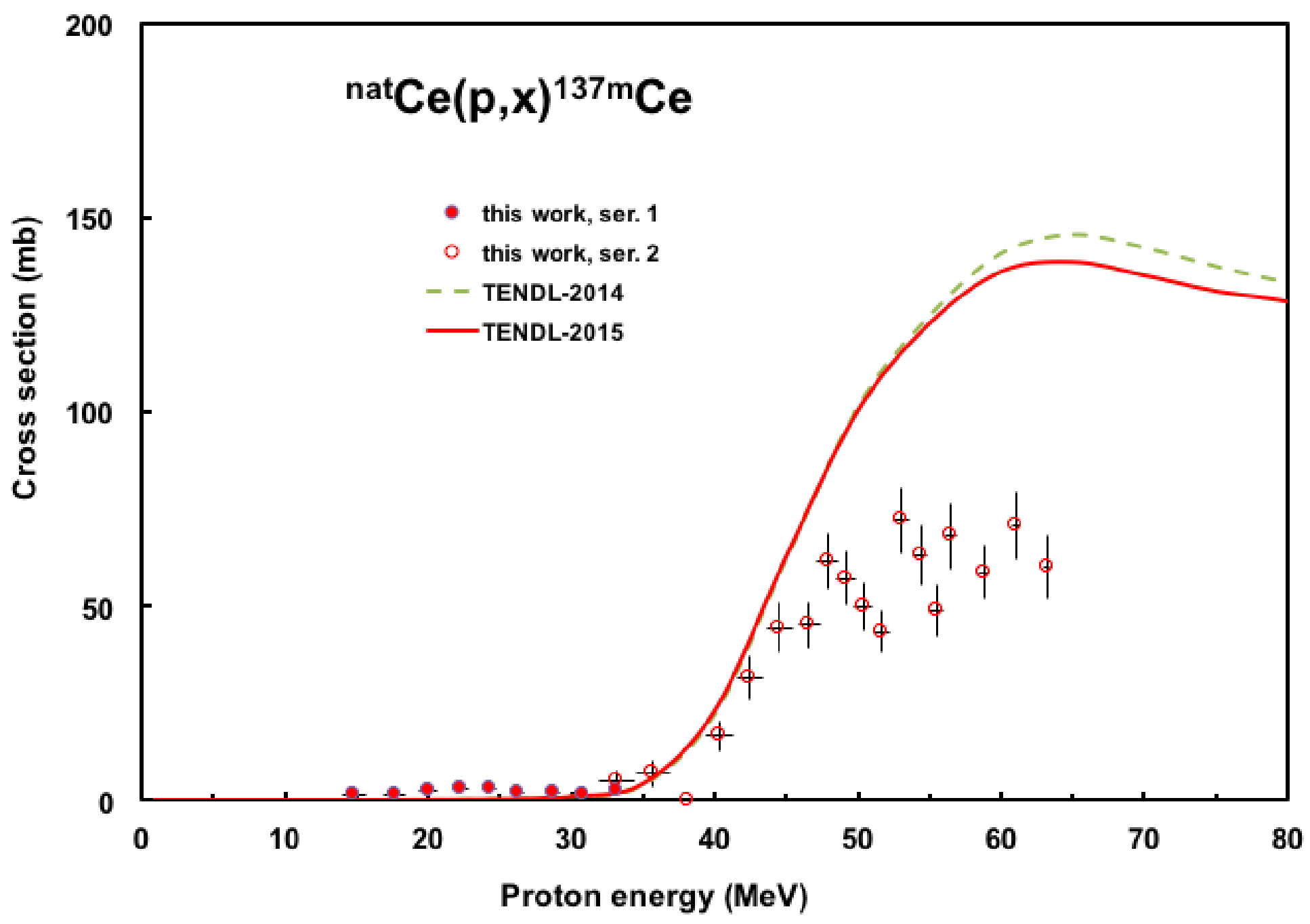

Fig. 8. Experimental and theoretical excitation functions for the $^{nat}Ce(p,x)^{137m}Ce$ reaction

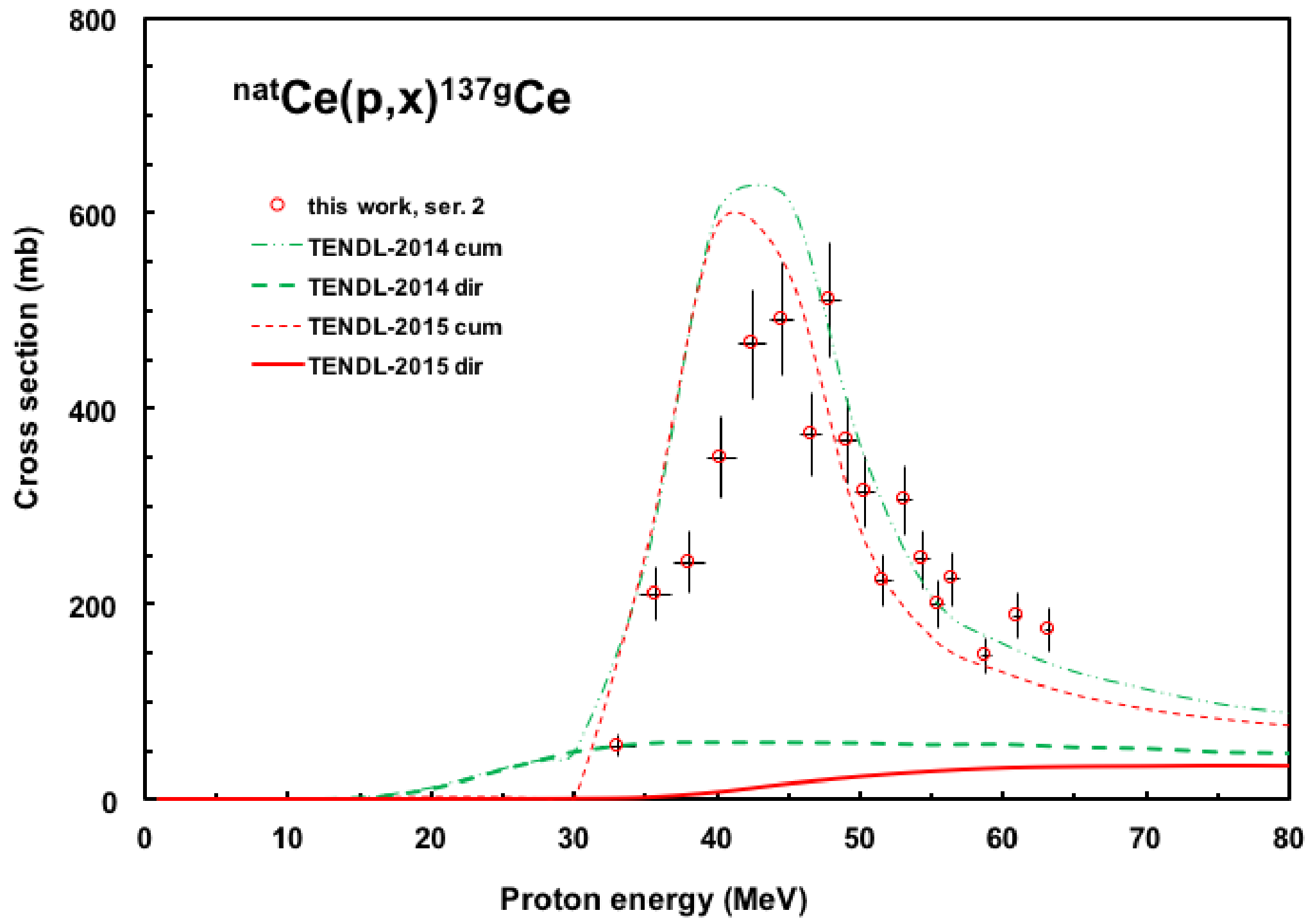

Fig. 9. Experimental and theoretical excitation functions for the $^{nat}Ce(p,x)^{137g}Ce$ reactions

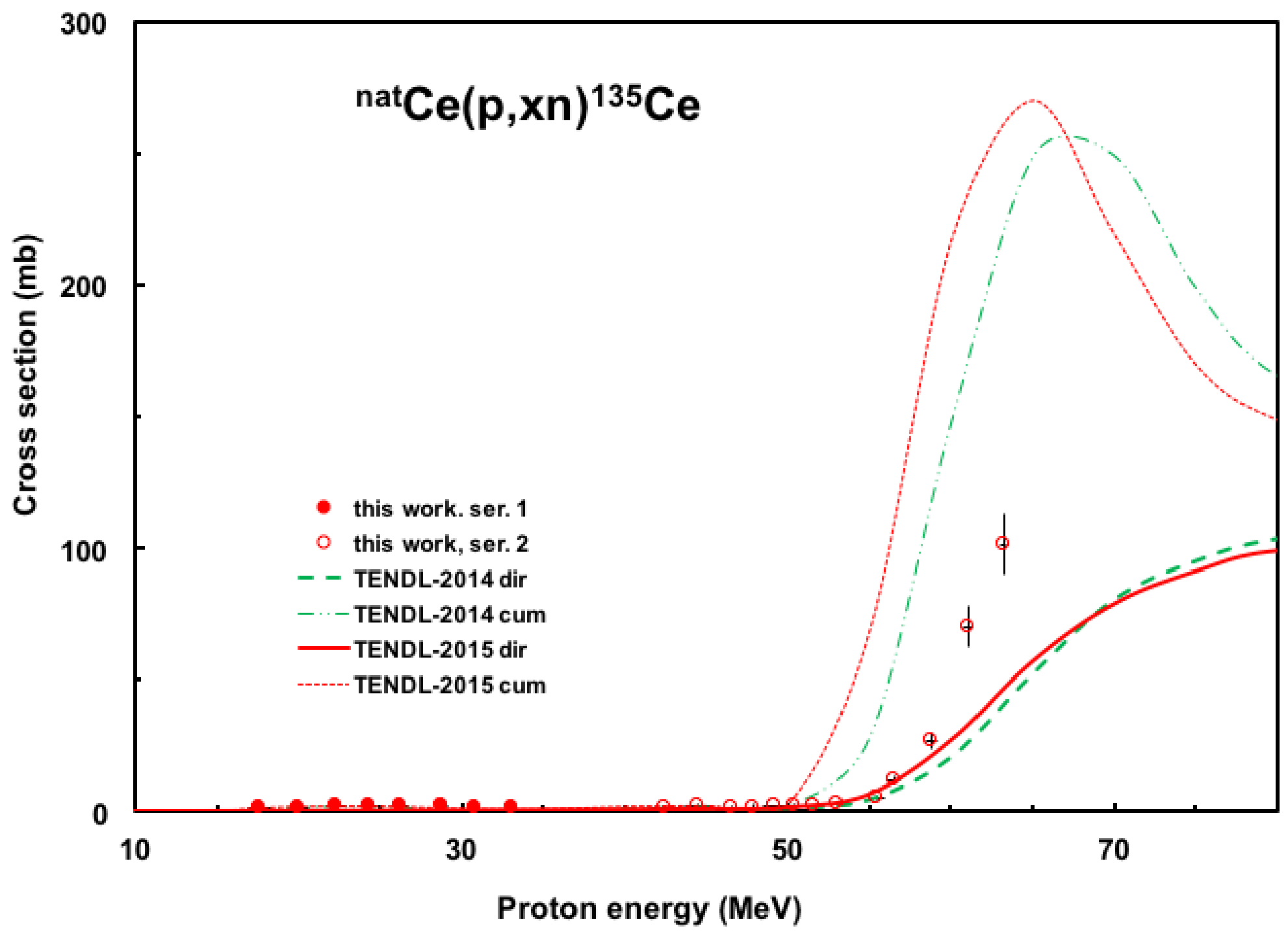

Fig. 10. Experimental and theoretical excitation functions for the $^{nat}Ce(p,x)^{135}Ce$ reaction

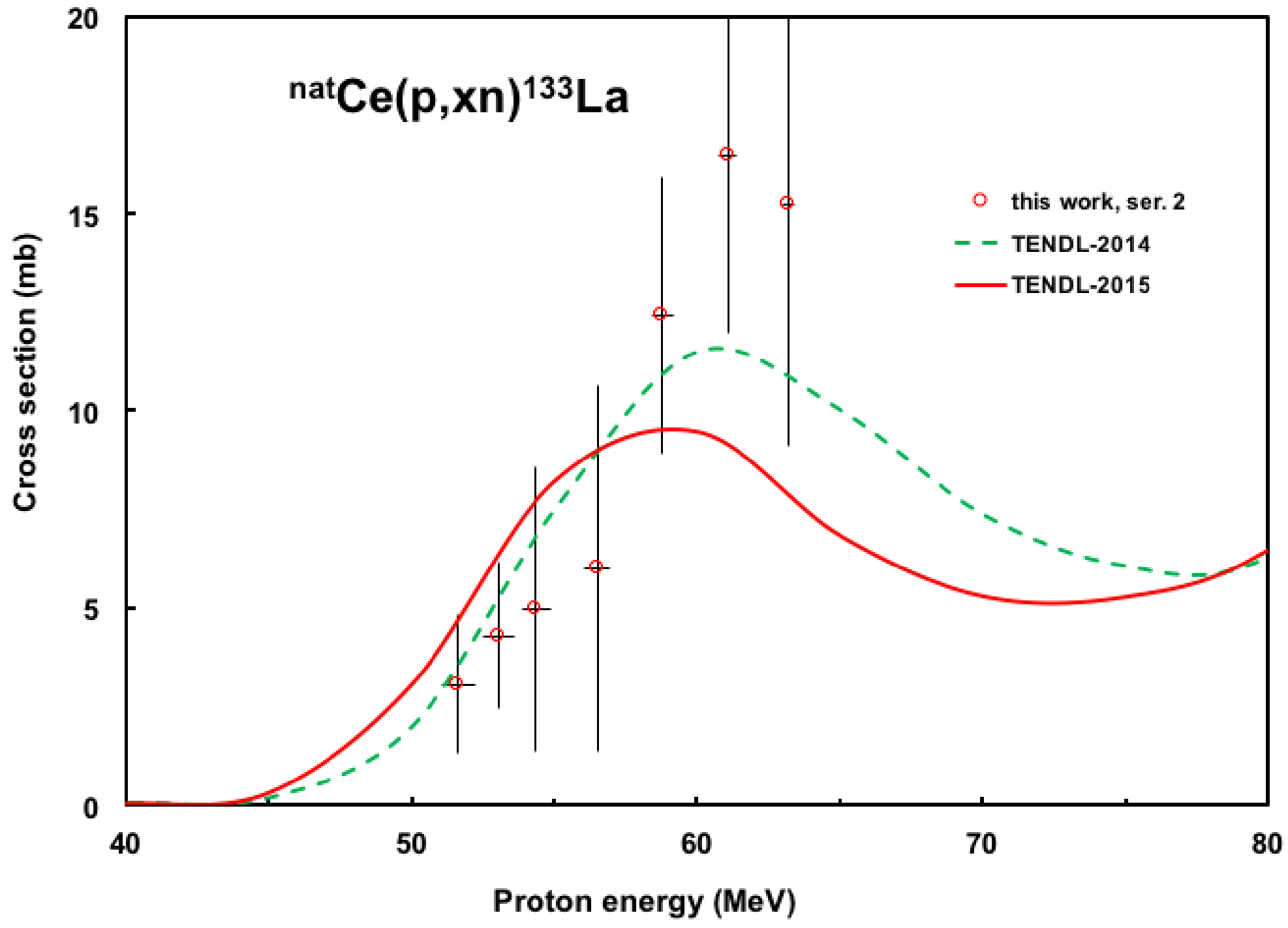

Fig. 11. Experimental and theoretical excitation functions for the $^{nat}Ce(p,x)^{133}La$ reaction

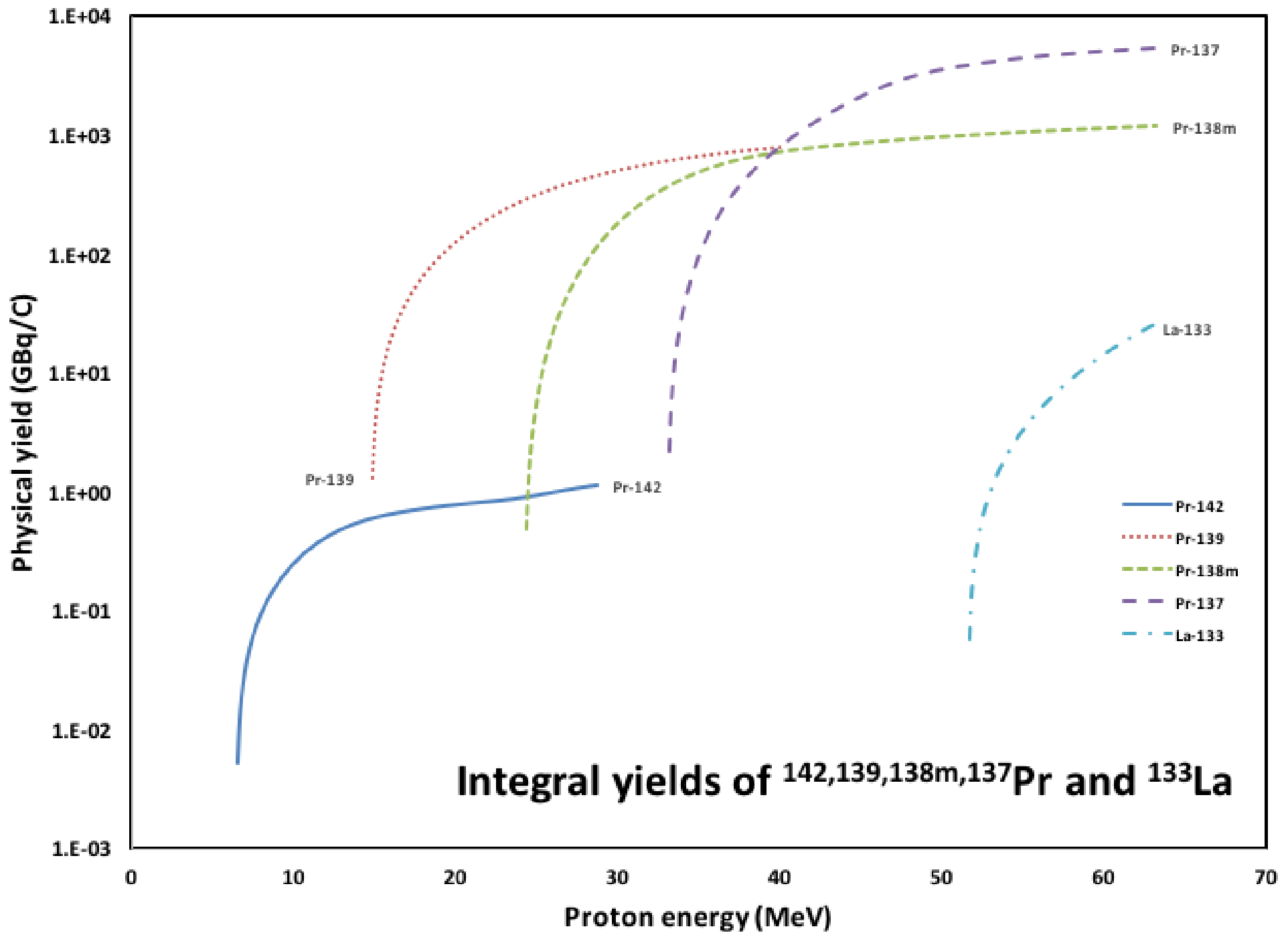

Fig. 12. Integral yields for production of $^{142,139,138m,137}Pr$ and $^{140}La$ deduced from the excitation functions

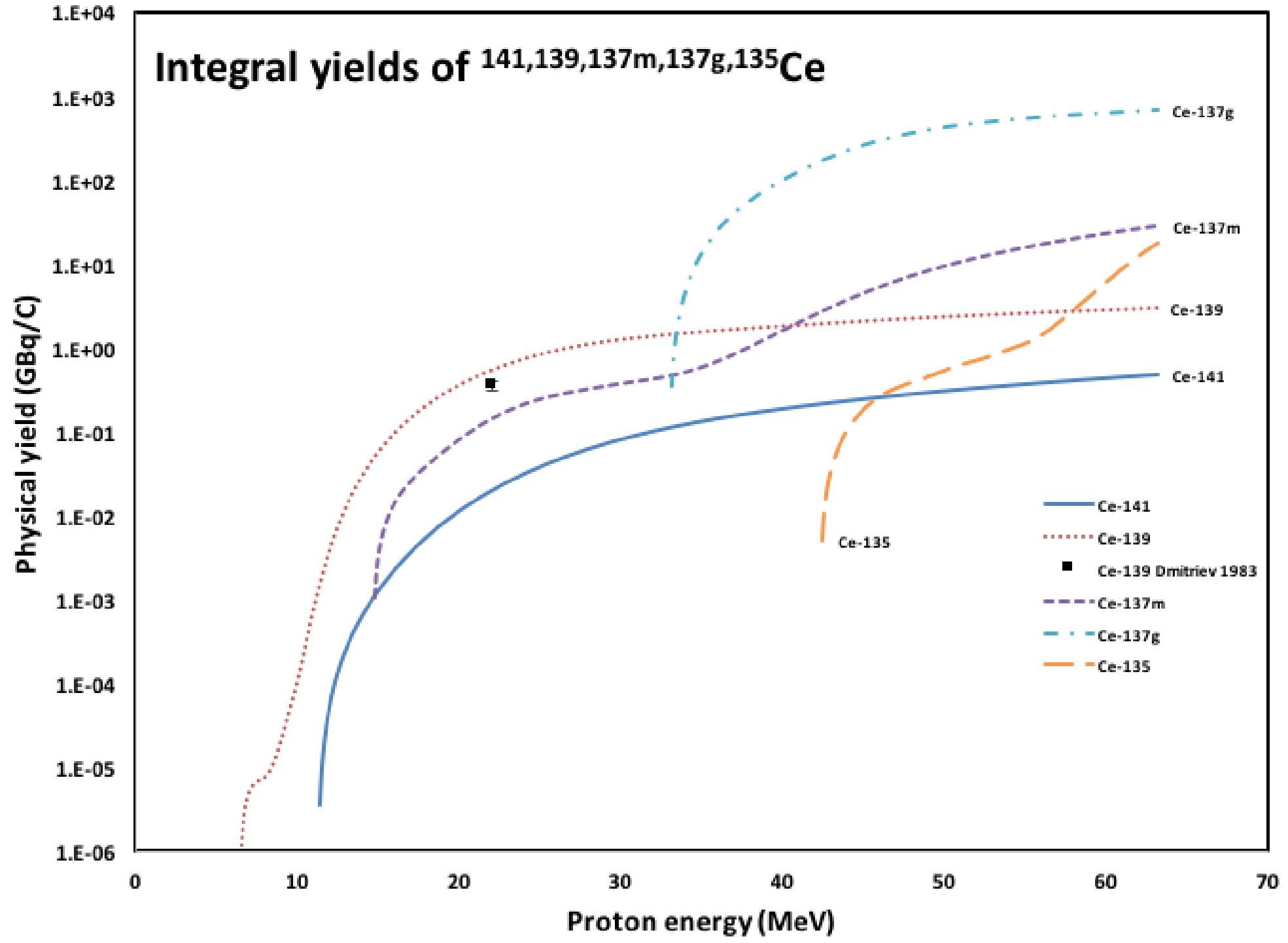

Fig. 13. Integral yields for production of $^{143,141,139,137m,137g,135}Ce$ deduced from the excitation functions in comparison with the literature experimental integral yields

## References

[1] M. Neves, A. Kling, A. Oliveira, Radionuclides used for therapy and suggestion for new candidates, J. Radioanal. Nucl. Chem., 266 (2005) 377-384.

[2] H. Uusijarvi, P. Bernhardt, F. Rosch, H.R. Maecke, E. Forssell-Aronsson, Electron- and positron-emitting radiolanthanides for therapy: Aspects of dosimetry and production, J Nucl Med, 47 (2006) 807-814.

[3] B. Thomadsen, W. Erwin, F. Mourtada, The Physics and Radiobiology of Targeted Radionuclide Therapy, in: T.W. Speer (Ed.) Targeted Radionuclide Therapy, Lippincott Williams & Wilkins, 2011, pp. 71.

[4] C. Vermeulen, G.F. Steyn, F.M. Nortier, F. Szelecsényi, Z. Kovács, S.M. Qaim, Production of $^{139}$Ce by proton-induced reactions on $^{141}$Pr and $^{nat}$La, Nucl. Instrum. Methods Phys. Res., Sect. B, 255 (2007) 331-337.

[5] G.F. Steyn, C. Vermeulen, F.M. Nortier, F. Szelecsenyi, Z. Kovacs, S.M. Qaim, Production of no-carrier-added Pr-139 via precursor decay in the proton bombardment of Pr-nat, Nucl. Instrum. Methods Phys. Res., Sect. B, 252 (2006) 149-159.

[6] M.K. Bakht, M. Sadeghi, C. Tenreiro, A novel technique for simultaneous diagnosis and radioprotection by radioactive cerium oxide nanoparticles: study of cyclotron production of Ce-137m, J. Radioanal. Nucl. Chem., 292 (2012) 53-59.

[7] S.W. Zielhuis, J.H. Seppenwoolde, V.A. Mateus, C.J. Bakker, G.C. Krijger, G. Storm, B.A. Zonnenberg, A.D. van het Schip, G.A. Koning, J.F. Nijsen, Lanthanide-loaded liposomes for multimodality imaging and therapy, Cancer Biother Radiopharm, 21 (2006) 520-527.

[8] F. Tárkányi, S. Takács, F. Ditrói, A. Hermanne, H. Yamazaki, M. Baba, A. Mohammadi, A.V. Ignatyuk, Activation cross-sections of deuteron induced nuclear reactions on neodymium up to 50 MeV, Nucl. Instrum. Methods Phys. Res., Sect. B, 325 (2014) 15-26.

[9] A. Hermanne, F. Tarkanyi, S. Takacs, F. Ditroi, Extension of excitation functions up to 50 MeV for activation products in deuteron irradiations of Pr and Tm targets, Nucl. Instrum. Methods Phys. Res., Sect. B, 383 (2016) 81-88.

[10] F. Tarkanyi, S. Takacs, F. Ditroi, J. Csikai, A. Hermanne, A.V. Ignatyuk, Activation cross-section measurement of deuteron induced reactions on cerium for biomedical applications and for development of reaction theory, Nucl. Instrum. Methods Phys. Res., Sect. B, 316 (2013) 22-32.

[11] F. Tarkanyi, A. Hermanne, F. Ditroi, S. Takacs, Activation cross section data of proton induced nuclear reactions on lanthanum in the 34-65 MeV energy range and application for production of medical radionuclides, J. Radioanal. Nucl. Chem., 312 (2017) 691-704.

[12] A. Hermanne, R. Adam-Rebeles, F. Tárkányi, S. Takács, J. Csikai, M.P. Takács, A.V. Ignatyuk, Deuteron induced reactions on Ho and La: experimental excitation functions and comparison with code results, Nucl. Instrum. Methods Phys. Res., Sect. B, 311 (2013) 102-111.

[13] R.R. Kinsey, C.L. Dunford, J.K. Tuli, T.W. Burrows, in Capture Gamma – Ray Spectroscopy and Related Topics, Vol. 2. (NUDAT 2.6 http://www.nndc.bnl.gov/nudat2/), Springer Hungarica Ltd, Budapest, 1997.

[14] F. Tárkányi, S. Takács, K. Gul, A. Hermanne, M.G. Mustafa, M. Nortier, P. Oblozinsky, S.M. Qaim, B. Scholten, Y.N. Shubin, Z. Youxiang, Beam monitor reactions (Chapter 4). Charged particle cross-section database for medical radioisotope production: diagnostic radioisotopes and monitor reactions. , in: TECDOC 1211, IAEA, 2001, pp. 49.

[15] B. Pritychenko, A. Sonzogni, Q-value calculator, in, NNDC, Brookhaven National Laboratory, 2003.

[16] H.H. Andersen, J.F. Ziegler, Hydrogen stopping powers and ranges in all elements. The stopping and ranges of ions in matter, Volume 3., Pergamon Press, New York, 1977.

[17] F. Tárkányi, F. Szelecsényi, S. Takács, Determination of effective bombarding energies and fluxes using improved stacked-foil technique, Acta Radiol., Suppl., 376 (1991) 72.

[18] International-Bureau-of-Weights-and-Measures, Guide to the expression of uncertainty in measurement, 1st ed., International Organization for Standardization, Genève, Switzerland, 1993.

[19] A.J. Koning, D. Rochman, S. van der Marck, J. Kopecky, J.C. Sublet, S. Pomp, H. Sjostrand, R. Forrest, E. Bauge, H. Henriksson, O. Cabellos, S. Goriely, J. Leppanen, H. Leeb, A. Plompen, R. Mills, TENDL-2014: TALYS-based evaluated nuclear data library, in, www.talys.eu/tendl2014.html, 2014.

[20] A.J. Koning, D. Rochman, J. Kopecky, J.C. Sublet, E. Bauge, S. Hilaire, P. Romain, B. Morillon, H. Duarte, S. van der Marck, S. Pomp, H. Sjostrand, R. Forrest, H. Henriksson, O. Cabellos, G. S., J. Leppanen, H. Leeb, A. Plompen, R. Mills, TENDL-2015: TALYS-based evaluated nuclear data library,, in, https://tendl.web.psi.ch/tendl_2015/tendl2015.html, 2015.

[21] A.J. Koning, D. Rochman, Modern Nuclear Data Evaluation with the TALYS Code System, Nucl. Data Sheets, 113 (2012) 2841.

[22] M. Bonardi, The contribution to nuclear data for biomedical radioisotope production from the Milan cyclotron facility, in: K. Okamoto (Ed.) Consultants Meeting on Data Requirements for Medical Radioisotope Production, IAEA, INDC(NDS)-195 (1988), Tokyo, Japan, 1987, pp. 98.

[23] P.P. Dmitriev, G.A. Molin, Radioactive nuclide yields for thick target at 22 MeV proton energy, Vop. At. Nauki i Tekhn., Ser.Yadernye Konstanty, 44 (1981) 43.